\documentclass[a4paper,12pt]{article}
\usepackage{natbib}
\usepackage[colorlinks,citecolor=blue,urlcolor=blue]{hyperref}
\usepackage{epsf}
\usepackage{epsfig}
\usepackage{multirow}
\usepackage[english]{babel}
\usepackage{graphicx}
\usepackage{geometry}
\usepackage{subfigure}
\usepackage{natbib}
\usepackage{amssymb}
\usepackage{amsmath}
\usepackage{color}
\usepackage{enumerate}
\usepackage{authblk}

\newcommand{\specificthanks}[1]{\@fnsymbol{#1}}

\newcommand{\PP}{\mathbb{P}}

\title{Model Based Clustering for Mixed Data: clustMD}
\author[]{Damien McParland\thanks{damien.mcparland@ucd.ie} }
\author[]{Isobel Claire Gormley\thanks{claire.gormley@ucd.ie}}
\affil[]{School of Mathematics and Statistics,\protect \\ University College Dublin,\protect \\ Ireland.}

\date{}

\begin{document}

\maketitle

\begin{abstract}
A model based clustering procedure for data of mixed type, clustMD, is developed using a latent variable model.  It is proposed that a latent variable, following a mixture of Gaussian distributions, generates the observed data of mixed type. The observed data may be any combination of continuous, binary, ordinal or nominal variables. 

clustMD employs a parsimonious covariance structure for the latent variables, leading to a suite of six clustering models that vary in complexity and provide an elegant and unified approach to clustering mixed data. 

An expectation maximisation (EM) algorithm is used to estimate clustMD; in the presence of nominal data a Monte Carlo EM algorithm is required. The clustMD model is illustrated by clustering simulated mixed type data and prostate cancer patients, on whom mixed data have been recorded.
\end{abstract}

\section{Introduction}
\label{sec:intro}

Clustering mixed data has received increasing attention in the literature for some time. Early work involved the use of mixture models \citep{everitt88a, muthen99}, with the location mixture model \citep{lawrence96, hunt99, willse99} providing an alternative approach. The use of copula models for clustering mixed data has also received very recent attention \citep{marbac14, kosmidis14}.

Latent factor models have also been employed to model mixed data. \cite{quinn04}, \cite{gruhl13} and \cite{murray13} use factor analytic models to analyse mixed data but not in a clustering context. Latent variable models for clustering categorical or mixed data have been proposed by \cite{cai11, browne12, morlini11, viroli12} and \cite{gollini13}. However none of these can analyze the combination of continuous, nominal and ordinal variables without transforming the original variables. The mixture of factor analysers model for mixed data \citep{mcparland13, mcparland14a, mcparland14b} is a finite mixture model based on a combination of factor models, item response theory models and ideas from the multinomial probit model, with clustering mixed data capabilities. While the MFA-MD model can explicitly model the inherent nature of each variable type directly, it can be computationally expensive.

The proposed clustMD model is a mixture of latent Gaussian distributions, and provides a parsimonious and computationally efficient approach to clustering mixed data. The mixture of Gaussian distributions has become a traditional approach to clustering continuous data and  parsimonious versions of these models were developed by \cite{banfield93} and \cite{celeux95}. \cite{fraley02} provide a detailed overview of such models which are efficiently implementable through the {\tt mclust} software \citep{fraley12}. A similar parsimony ethos underpins the proposed clustMD model and a suite of models of varying levels of parsimony are developed.

An expectation maximisation (EM) algorithm \citep{dempster77} is used for inference. In the presence of nominal data, the expectation step is intractable so a Monte Carlo EM algorithm is required. The performance of clustMD is demonstrated by a simulation study and by clustering a group of prostate cancer patients, on whom variables of mixed type have been recorded.

\section{The clustMD model}
\label{sec:Model}

Mixture models are a very useful clustering tool and much research has been devoted to their development. Finite mixture models assume the data arise from a finite number of homogeneous clusters. Detailed explanations of mixture models may be found in \cite{titterington85, mclachlan00} and \cite{fruhwirth06}. The clustMD model employs a mixture of latent variable models to cluster mixed type data. In brief, the clustMD model assumes the observed $J$ mixed type variables in each observation vector $\underline{y}_i$ are a manifestation of an underlying latent continuous vector, $\underline{z}_{i}$ (for $i = 1, \ldots, N$), which follows a Gaussian mixture distribution.

\subsection{Modelling continuous data}
\label{subsec:Cns}

Under the clustMD model, continuous variables follow a multivariate Gaussian distribution, i.e. if variable $j$ is continuous, $y_{ij}  = z_{ij} \sim \mbox{N}(\mu_j, \sigma_j^2)$.

\subsection{Modelling ordinal data}
\label{subsec:Ord}

In the case of an ordinal variable, it is supposed that the observed response, $y_{ij}$ is a categorical manifestation of the latent continuous variable, $z_{ij}$, as is typical  in item response theory models \citep{johnson99, fox10}. 

For ordinal variable $j$ with $K_j$ levels let $\underline{\gamma}_{j}$ denote a $K_j+1$ vector of thresholds that partition the real line. The value of the latent $z_{ij}$ in relation to $\underline{\gamma}_{j}$ determines the observed ordinal response $y_{ij}$. The threshold parameters are constrained such that $- \infty = \gamma_{j,0} \leq \gamma_{j,1} \leq \ldots \leq \gamma_{j,K_j} = \infty$. If the latent $z_{ij}$ is such that $\gamma_{j,k-1} < z_{ij} < \gamma_{j, k}$ then the observed ordinal response, $y_{ij} = k$. The latent $z_{ij}$ follows a Gaussian distribution i.e. $z_{ij} \sim \mbox{N}(\mu_{j}, \sigma_{j}^2)$. Thus the probability of observing level $k$ can be expressed as the difference between two Gaussian cumulative distribution functions (CDF) denoted by $\Phi$:
$\PP(y_{ij}=k) = \Phi\left(\frac{\gamma_{j,k} - \mu_j}{\sigma_j}\right) - \Phi\left(\frac{\gamma_{j,k-1} - \mu_j}{\sigma_j}\right)$

The threshold parameters are invariant under translation and their values are not of primary interest in clustMD. Thus, for reasons of identifiability and efficiency, $\gamma_{j, k}$ is fixed such that $\gamma_{j, k} = \Phi^{-1}(\delta_{k})$, where $\delta_{k}$ is the proportion of the observed values of variable $J$ which are less than or equal to level $k$.

A binary variable can be thought of as an ordinal variable with two levels, denoted 1 and 2. Thus if variable $j$ is binary, then $\PP(y_{ij}=2) = 1 - \Phi\left(\frac{\gamma_{j,1} - \mu_j}{\sigma_j}\right)$.

\subsection{Modelling nominal data}
\label{subsec:Nom}

Nominal variables are more difficult to model since the set of possible responses is unordered. In this case, a multivariate latent vector is assumed to underlie the observed nominal variable. For nominal variable $j$ with $K_j$ possible responses, the underlying continuous vector has $K_j-1$ dimensions, i.e. $\underline{z}_{ij} = (z_{ij}^1, \ldots, z_{ij}^{K_j - 1}) \sim \mbox{MVN}_{K_{j}-1}(\underline{\mu}_j, \Sigma_j)$. The observed nominal response $y_{ij}$ is a manifestation of the values of the elements of $\underline{z}_{ij}$ relative to each other and to a threshold, assumed to be $0$. That is,
$$y_{ij} = \left\{ \begin{array}{ll}
         1 & \mbox{if $ \displaystyle \max_k\{z_{ij}^k\} < 0$}; \vspace{0.5cm}\\ 
         k & \mbox{if $z_{ij}^{k-1} = \displaystyle \max_k\{z_{ij}^k\}$ and $z_{ij}^{k-1} > 0 \:\:$ for $k = 2, \ldots, K_{j}$}.\\\end{array} \right. $$
Binary data can be considered as nominal with two unordered responses. This model for nominal data is equivalent to the proposed ordinal data model in such a case. A similar latent variable approach to modelling nominal data is the mutinomial probit model \citep{geweke94}.

\subsection{A joint model for mixed data}
\label{subsec:Mix}

Let $Y$ denote a data matrix with $N$ rows and $J$ columns. Without loss of generality, suppose that the continuous variables are in the first $C$ columns, the ordinal and binary variables are in the following $O$ columns and the nominal variables are in the final $J-(C+O)$ columns. The latent continuous data underlying both the ordinal and nominal data are assumed to be Gaussian, as are any observed continuous data. Thus the joint vector of observed and latent continuous data is assumed to follow a multivariate Gaussian distribution $\underline{z}_{i} \sim \mbox{MVN}_P(\underline{\mu}, \Sigma)$.
Since more than one latent dimension is required to model each nominal variable $P = C + O + \sum_{j=C+O+1}^{J}(K_j -1)$. This model provides a unified way to simultaneously model continuous, ordinal and nominal data.

\subsection{A mixture model for mixed data}
\label{subsec:Mixture}

The joint model for mixed data is embedded in a finite mixture model, facilitating the clustering of mixed data. This model, clustMD, is closely related to the parsimonious mixture of Gaussian distributions \citep{banfield93, celeux95}.  In clustMD, it is assumed that $\underline{z}_i$ follows a mixture of $G$ Gaussian distributions i.e. $\underline{z}_i \sim \sum_{g=1}^{G} \pi_g \mbox{MVN}_{P}(\underline{\mu}_g, \Sigma_g)$ where $\pi_g$ is the marginal probability of belonging to cluster $g$ and $\underline{\mu}_g$ and $\Sigma_g$  denote the mean and covariance for cluster $g$ respectively.

\subsection{Decomposing the covariance matrix}
\label{subsec:matrixdecomp}

Gaussian parsimonious mixture models utilise an eigenvalue decomposition of the cluster covariance matrix $\Sigma_g = \lambda_gD_gA_gD_g$ where $|A_g|=1$. The $\lambda_g$ parameter controls the cluster volume, $D_g$ is a matrix of eigenvectors of $\Sigma_g$ that controls the orientation of the cluster and $A_g$ is a diagonal matrix of eigenvalues of $\Sigma_g$ that controls the shape of the cluster. The decomposed covariance is constrained in various ways to produce parsimonious models. 

The covariance matrix for the clustMD model is assumed to be diagonal, meaning that $D_g = I$, the identity matrix. This assumption imples that variables are conditionally independent given their cluster membership. Thus under clustMD $\Sigma_g = \lambda_g A_g$. These parameters can then be constrained to be different or equal across groups and $A$ can also be constrained to be the identity matrix. This gives rise to a suite of  6 clustMD models with varying levels of parsimony. The 6 clustMD models and corresponding constraints are detailed in Table~\ref{tab:mods}.

\begin{table}
\caption{Covariance matrix structures of varying degrees of parsimony. Parameters are unconstrained (U), constrained (C) to be equal across groups or equal to the identity (I).}
\label{tab:mods}
\begin{tabular}{l | l | l | l | l | l}
\hline\noalign{\smallskip}
&&&&\multicolumn{2}{c}{\# Covariance parameters} \\
Model & $\lambda$ & $A$ & $D$ & No nominal variables  & Nominal variables\\
\noalign{\smallskip}\hline\noalign{\smallskip}
 $EII$ & C & I & I& 1 & 1 \\
 $VII$ & U & I & I& $G$ & $2G - 1$ \\
 $EEI$ & C & C & I& $1 + P$ & $C+O$ \\
 $VEI$ & U & C & I& $G + P$ & $2G + C + O - 2$\\
 $EVI$ & C & U & I& $1 + GP$ & $G(P-2) + C + O - P + 2$\\
 $VVI$ & U & U & I& $G(1+P)$ & $P(G-1) + O$\\
\noalign{\smallskip}\hline
\end{tabular}
\end{table}

\subsubsection{Identifying clustMD in the presence of nominal variables}

If no nominal variables are present in the data, then the clustMD model is identified because the threshold parameters are fixed. However, in the presence of nominal variables, the model as it stands is not identified. Infinitely many combinations of the model parameters give rise to the same likelihood. Constraints must be placed on the parameters relating to nominal variables in order to obtain consistent parameter estimates. As in \cite{viroli12}, the constraint $\sum_g \pi_g \mu_{gp} = 0$ for each dimension $p$ corresponding to a nominal variable is applied across the suite of models, which amounts to insisting that $\mathbb{E}(z_{ip}) = 0$ for $p = C+O+1, \ldots, P$.  Further, a separate volume parameter $\tilde{\lambda}_g$ which applies only to the latent dimensions corresponding to nominal variables is also required. The diagonal elements of $\Sigma_g$ corresponding to these dimensions are $\tilde{\lambda}_g a_{gp}$, where $a_{gp}$ is the $p^{th}$ diagonal element of $A_g$. 

Different constraints on $\tilde{\lambda}_g$ are required in the different clustMD models. For example, the $EII$ model is identified by fixing $\tilde{\lambda}=1$, meaning that the diagonal elements of $\Sigma$ corresponding to nominal variables are simply set to 1. The $VII$ model is identified by insisting that $\sum_g \tilde{\lambda}_g = 1$. This may be accomplished by dividing each $\tilde{\lambda}_g$ by $\sum_g \tilde{\lambda}_g$ after each iteration of the model fitting algorithm. To identify the $EEI$ model $\tilde{\lambda}$ is set to 1, as is $a_{p}$ for $p$ corresponding to nominal variables. The $VEI$ model is constrained so that  $\sum_g \tilde{\lambda}_g = 1$ and $a_{p} = 1$ for nominal dimensions $p$. Thus the nominal portions of the $EEI$ and $VEI$ models are the same as the $EII$ and $VII$ models respectively. 

The $EVI$ model is identified by fixing $\tilde{\lambda}=1$ and constraining $a_{gp}$ so that $\sum_g a_{gp}=1$ for nominal dimensions $p$. This constraint on $a_{gp}$ is implemented by dividing each $a_{gp}$ term by $\sum_g a_{gp}$ after each iteration of the model fitting algorithm. Finally the $VVI$ model is identified by constraining $\tilde{\lambda}_g$ and $a_{gp}$ so that $\sum_g \tilde{\lambda}_g = 1$ and $\sum_g a_{gp}=1$ for each nominal dimension $p$. It is possible to fit all 6 clustMD models, even in the presence of nominal data. However, there are, in reality, only 4 models for the nominal portion of the clustMD model. 

\section{Fitting the clustMD model}
\label{sec:modelfit}


The clustMD model is fitted using an EM algorithm. If nominal data are present then a Monte Carlo approximation is required for the expectation step and hence the algorithm is a Monte Carlo EM (MCEM) algorithm. 

\subsection{Deriving the complete data log likelihood}
\label{subsec:loglikelihood}

The categorical part of each observation can be thought of as one of a (possibly large) number, $M$, of response patterns. Let $\underline{y}_i^\beta$ be a binary vector of length $M$ indicating which response pattern is observed, i.e. if response pattern $m$ is observed $y_{im} = 1$; all other entries are 0. Thus, $\underline{y}_i^\beta \sim \mbox{Multinomial}(1, \underline{q})$ where $\underline{q} = (q_1, \ldots, q_M)$ and $q_m = \int_{\Omega_m} f(\underline{z}_i) d\underline{z}_i$. The portion of $\mathbb{R}^{P-C}$ that generates pattern $m$ is denoted $\Omega_m$.  Let $\underline{z}_i^\beta$ denote the latent continuous vector corresponding to the observed categorical variables and the superscript $\beta$ denote the portions of the model parameters corresponding to these data. A binary latent variable, $\underline{\ell}_{i}$ is introduced that indicates the cluster membership of observation $i$, i.e. $\ell_{ig}=1$ if observation $i$ belongs to cluster $g$; all other entries are 0. Thus the joint density of $\underline{z}_i^\beta$, $\underline{y}_i^\beta$ and $\underline{\ell}_i$ can be written as
$$f(\underline{z}_i^\beta, \underline{y}_i^\beta, \underline{\ell}_i) = f(\underline{z}_i^\beta | \underline{y}_i^\beta, \ell_{ig}=1)f(\underline{y}_i^\beta | \ell_{ig} = 1)f(\underline{\ell}_i)$$
where:
\begin{itemize}
 \item $\underline{\ell}_i \sim \mbox{Multinomial}(1, \underline{\pi})$
 \item $\underline{y}_i^\beta | \ell_{ig} = 1 \sim \mbox{Multinomial}(1, \underline{q}_g)$ where $\underline{q}_g = (q_{g1}, \ldots, q_{gM})\\ \mbox{ and } q_{gm} = \int_{\Omega_m} f(\underline{z}_i^\beta | \ell_{ig} = 1) d\underline{z}_i^\beta = \int_{\Omega_m} \mbox{MVN}(\underline{z}_i^\beta | \underline{\mu}_g^\beta, \Sigma_g^\beta) d\underline{z}_i^\beta$
 \item $\underline{z}_i^\beta | \underline{y}_i^\beta, \ell_{ig}=1 \sim \mbox{MVN}^T(\underline{z}_i^\beta | \underline{\mu}_g^\beta, \Sigma_g^\beta)$, a truncated multivariate Gaussian distribution. The points of truncation are those which satisfy the ordinal and/or nominal conditions detailed in Sections \ref{subsec:Ord} and \ref{subsec:Nom} given $\underline{y}_i^\beta$.
\end{itemize}
Thus,
 \begin{eqnarray*}
f(\underline{z}_i^\beta, \underline{y}_i^\beta, \underline{\ell}_i) &\propto& \left\{\prod_{g=1}^{G}\left[\frac{\mbox{MVN}(\underline{z}_i^\beta | \underline{\mu}_g^\beta, \Sigma_g^\beta)}{\prod_{m=1}^{M} q_{gm}^{y_{im}^\beta}} \right]^{\ell_{ig}} \right\}  \: \left\{\prod_{g=1}^{G}\prod_{m=1}^{M} \left[ q_{gm}^{y_{im}^\beta} \right]^{\ell_{ig}} \right\} \: \left\{\prod_{g=1}^{G}\pi_g^{\ell_{ig}}\right\}\\
&=& \prod_{g=1}^{G}\left[ \pi_g  \mbox{MVN}(\underline{z}_i^\beta | \underline{\mu}_g^\beta, \Sigma_g^\beta) \right]^{\ell_{ig}}.
\end{eqnarray*}
Let $\underline{y}_{i}^\alpha = \underline{z}_i^\alpha$ denote the observed continuous variables and the superscript $\alpha$ denote the portions of the model parameters that apply to continuous variables. Since $\Sigma_g$ is assumed to be diagonal, the complete data likelihood is the product of the likelihood of the continuous variables and the likelihood of the latent variables relating to the observed categorical variables:
\begin{eqnarray}
\mathcal{L}_c &=& \prod_{i=1}^{N} \prod_{g=1}^{G} \left[ \pi_g  \mbox{MVN}(\underline{z}_i^{\alpha} | \underline{\mu}_g^{\alpha}, \Sigma_g^{\alpha}) \times  \mbox{MVN}(\underline{z}_i^\beta | \underline{\mu}_g^\beta, \Sigma_g^\beta) \right]^{\ell_{ig}} \label{eqn:likelihood}
\\
\Rightarrow \log \mathcal{L}_c & = & \sum_{i=1}^{N} \sum_{g=1}^{G}  \left[ \ell_{ig} \log\pi_g + B - \frac{\ell_{ig}}{2} \log|\Sigma_g| - \frac{\ell_{ig}}{2}\left( \underline{z}_i^{\alpha^T} \Sigma_g^{\alpha^{-1}}\underline{z}_i^{\alpha} +  \underline{z}_i^{\beta^T} \Sigma_g^{\beta^{-1}}\underline{z}_i^{\beta} \right) \right. \nonumber \\
  && \left. + \ell_{ig}\left(  \underline{\mu}^{\alpha^T} \Sigma_g^{\alpha^{-1}}\underline{z}_i^{\alpha}
  +  \underline{\mu}^{\beta^T} \Sigma_g^{\beta^{-1}}\underline{z}_i^{\beta} \right) - \frac{\ell_{ig}}{2} \left( \underline{\mu}^{\alpha^T} \Sigma_g^{\alpha^{-1}}\underline{\mu}^{\alpha}
  +  \underline{\mu}^{\beta^T} \Sigma_g^{\beta^{-1}}\underline{\mu}^{\beta} \right) \right] \nonumber
\end{eqnarray}
where $B$ denotes a constant.

\subsubsection{The expectation step}
\label{subsubsec:estep}
The expectation step (E-step) of the EM algorithm consists of computing the expectation of the complete log likelihood with respect to the latent data $\underline{z}_i^\beta$ and the latent cluster labels $\ell_{ig}$. Three expectations $\mathbb{E}(\ell_{ig}|\underline{y}_i, \underline{\mu}_g, \Sigma_g, \pi_g)$, $\mathbb{E}(\ell_{ig}\underline{z}_i^\beta|\underline{y}_i, \underline{\mu}_g^\beta, \Sigma_g^\beta, \pi_g)$ and $\mathbb{E}(\ell_{ig}\underline{z}_i^{\beta^T}\underline{z}_i^\beta|\underline{y}_i, \underline{\mu}_g^\beta, \Sigma_g^\beta, \pi_g)$ are therefore required.

For the first expectation, since $\ell_{ig}$ takes the values 0 or 1, then:
\begin{eqnarray}
 \hspace{-0.4cm}\mathbb{E}(\ell_{ig}| \ldots) &=& \frac{\pi_g \mbox{MVN}(\underline{z}_i^\alpha|\underline{\mu}_g^\alpha, \Sigma_g^\alpha)\int_{\Omega_m}\mbox{MVN}(\underline{z}_i^\beta|\underline{\mu}_g^\beta, \Sigma_g^\beta)d\underline{z}_i^\beta}{\sum_{g'=1}^{G}\pi_{g'} \mbox{MVN}(\underline{z}_i^\alpha|\underline{\mu}_{g'}^\alpha, \Sigma_{g'}^\alpha)\int_{\Omega_m}\mbox{MVN}(\underline{z}_i^\beta|\underline{\mu}_{g'}^\beta, \Sigma_{g'}^\beta)d\underline{z}_i^\beta} = \tau_{ig} \label{eqn:e1}
\end{eqnarray}
Since the covariance matrix $\Sigma_g^\beta$ is assumed to be diagonal, the integrals in (\ref{eqn:e1}) can be expressed as a product of probabilities. The probabilities corresponding to ordinal variables are easily approximated given the threshold parameters. 

However, in the presence of nominal variables, calculating the probabilities is more challenging, due to the way in which the latent data generate a nominal response, detailed in Section \ref{subsec:Nom}. Thus, for each cluster, a Monte Carlo approximation of the probability of each possible response is obtained by simulating a large number of continuous vectors from a multivariate Gaussian distribution with mean $\underline{\mu}_g^j$ and covariance $\Sigma_g^j$, where $\underline{\mu}_g^j$ and $\Sigma_g^j$ are the portions of the mean vector and covariance matrix for cluster $g$, corresponding to nominal variable $j$. The probability of each response is approximated by the proportion of these simulations that generate each response. The Monte Carlo approximations can then be used to estimate $\tau_{ig}$ above.

Similar to \cite{karlis09}, the second expectation is
$$ \mathbb{E}(\ell_{ig}\underline{z}_i^\beta|\underline{y}_i^\beta, \underline{\mu}_g^\beta, \Sigma_g^\beta, \pi_g) = \PP(\ell_{ig}=1| \underline{\mu}_g^\beta, \Sigma_g^\beta, \pi_g) \mathbb{E}(\underline{z}_i^\beta|\ell_{ig}=1, \underline{\mu}_g^\beta, \Sigma_g^\beta, \pi_g) = \tau_{ig}\underline{m}_{ig}$$
and the third expectation is
\begin{eqnarray*}
 \mathbb{E}(\ell_{ig}\underline{z}_i^{\beta^T}\underline{z}_i^\beta| \underline{\mu}_g^\beta, \Sigma_g^\beta, \pi_g) &=& \PP(\ell_{ig}=1| \underline{\mu}_g^\beta, \Sigma_g^\beta, \pi_g)\mathbb{E}(\underline{z}_i^{\beta^T}\underline{z}_i^\beta| \ell_{ig}=1, \underline{\mu}_g^\beta, \Sigma_g^\beta, \pi_g)\\
 &=& \tau_{ig}\sum_p \mathbb{E}(z_{ip}^2|\ell_{ig}=1, \ldots) = \tau_{ig} \sum_p s_{igp}.
\end{eqnarray*}
The computation of $m_{igp}$ and $s_{igp}$ corresponding to ordinal variables is straightforward: given the relevant threshold parameters, they are simply the first and second moments of a truncated Gaussian distribution. In the case of dimensions relating to nominal variables, $m_{igp}$ and $s_{igp}$ are also related to the first and second moments of a truncated multivariate Gaussian, but they are difficult to calculate given the truncations outlined in Section \ref{subsec:Nom}. A Monte Carlo approximation again is used in these cases. Suppose that $y_{ij} = k$ for nominal variable $j$, then $\mathbb{E}(\underline{z}_i^j|y_{ij} = k, \ell_{ig}=1, \underline{\mu}_g, \Sigma_g, \pi_g)$ and $\mathbb{E}(\underline{z}_i^{j^T}\underline{z}_i^j|y_{ij} = k, \ell_{ig}=1, \underline{\mu}_g, \Sigma_g, \pi_g)$ must be calculated. The Monte Carlo samples generated to calculate the probabilities $\tau_{ig}$ for the first expectation can be reused to this end. For each possible response $k$ and each cluster $g$ the first moment can be approximated by calculating the sample mean of those Monte Carlo samples which generate response $k$. The second moment can be approximated by calculating the inner product of the vectors that generate response $k$ and then calculating the sample mean of these inner products. The second expectation can then be approximated by summing the elements of this sample mean vector.

\subsubsection{The maximisation step}
\label{subsubsec:mstep}

The maximisation (M-step) of the algorithm maximises the expected value, $Q$, of the complete log likelihood based on the current values of the model parameters. The M-step in the case of the $VVI$ model is derived below, other model derivations are provided in \cite{mcparland15}. The $VVI$ model is the most general of the 6 clustMD models i.e. $\Sigma_g = \lambda_g A_g$. The M-step maximises
\begin{eqnarray*}
  Q &=& \sum_g \log \pi_g \sum_i \tau_{ig} - \frac{C+O}{2}\sum_g\log\lambda_g\sum_i \tau_{ig} - \frac{P-C-O}{2}\sum_g\log\tilde{\lambda}_g\sum_i \tau_{ig} \\
 &&- \frac{1}{2}\sum_g\sum_p\log a_{gp}\sum_i\tau_{ig} - \frac{1}{2}\sum_g \sum_{p=1}^{C} \sum_i\frac{z_{ip}^2\tau_{ig}}{\lambda_g a_{gp}} - \frac{1}{2}\sum_g \sum_{p=C+1}^{C+O}\sum_i\frac{s_{igp}\tau_{ig}}{\lambda_g a_{gp}}\\
 &&- \frac{1}{2}\sum_g \sum_{p=C+O+1}^{P}\sum_i\frac{s_{igp}\tau_{ig}}{\tilde{\lambda}_g a_{gp}} + \sum_g\sum_{p=1}^{C+O}\sum_i\frac{\mu_{gp}z_{igp}^*\tau_{ig}}{\lambda_ga_{gp}}  + \sum_g\sum_{p=C+O+1}^P \sum_i\frac{\mu_{gp}z_{igp}^*\tau_{ig}}{\tilde{\lambda}_g a_{gp}} \\
 && - \frac{1}{2}\sum_g\sum_{p=1}^{C+O} \sum_i\frac{\mu_{gp}^2\tau_{ig}}{\lambda_g a_{gp}} - \frac{1}{2}\sum_g\sum_{p=C+O+1}^P \sum_i\frac{\mu_{gp}^2\tau_{ig}}{\tilde{\lambda}_g a_{gp}} + R
\end{eqnarray*}
where $R$ denotes a constant and $\underline{z}_{ig}^* = (\underline{z}_i^\alpha, \underline{m}_{ig})^T$. Maximising $Q$ with respect to $\lambda_g$ yields
$$\hat{\lambda}_g = \frac{ \sum_{p=1}^C \sum_i \frac{z_{ip}^2\tau_{ig}}{a_{gp}}  + \sum_{p=C+1}^{C+O} \sum_i \frac{s_{igp}\tau_{ig}}{a_{gp}} -\sum_{p=1}^{C+O}\frac{\mu_{gp}}{a_{gp}}\left[2\sum_i z_{igp}^*\tau_{ig} -\mu_{gp}\sum_i \tau_{ig}\right] }{ (C+O)\sum_i\tau_{ig}}$$
and, if nominal variables are present, maximising $Q$ with respect to $\tilde{\lambda}_g$ yields
$$\hat{\tilde{\lambda}}_g = \frac{  \sum_{p=C+O+1}^P \sum_i \frac{s_{igp}\tau_{ig}}{a_{gp}} -2\sum_{p=C+O+1}^P\frac{\mu_{gp}}{a_{gp}}\sum_i z_{igp}^*\tau_{ig} + \sum_{p=C+O+1}^P \frac{\mu_{gp}^2}{a_{gp}}\sum_i \tau_{ig} }{ (P-C-O)\sum_i\tau_{ig}}.$$
Maximising $Q$ with respect to $a_{gp}$ yields
$$\hat{a}_{gp} = \frac{\zeta_{gp} - 2\mu_{gp}\sum_i z_{igp}^*\tau_{ig} + \mu_{gp}^2\sum_i\tau_{ig}}{\lambda_g \xi_g \sum_i\tau_{ig}}$$
where $\lambda_g = \tilde{\lambda}_g$ if $p > (C+O)$, $\zeta_{gp}=\sum_i z_{ip}^2\tau_{ig}$ if $p \leq C$ and $\zeta_{gp}=\sum_i s_{igp} \tau_{ig}$ if $p > C$, $\xi_{g}=(\prod_{p=1}^{C+O} a_{gp})^{\frac{1}{C+O}}$ if $p \leq C+O$ and $\xi_{g}=1$ if $p > C+O$.

The (Monte Carlo) E and M steps are iterated until convergence is reached. Convergence is guaranteed even though a Monte Carlo approximation is used. However, the monotone increase in the likelihood at each iteration, which a standard EM algorithm guarantees, does not apply here. The example of \cite{wei90} is followed and the algorithm is terminated when a plot of the parameter estimates against the iteration number show that the process has stabilised. For more detail on convergence and the Monte Carlo EM algorithm see \cite{mclachlan08}.

The algorithm is initialised by obtaining an initial clustering and estimating model parameters based on that clustering. To avoid local minima a number of different intitialisations are used; namely $K$ means, hierarchical and random clustering. The sensitivity of the EM algorithm to initialising values is a known problem. Recent work on this issue includes that of \cite{ohagan12a}. However, for the data sets analysed in this paper, the (MC)EM algorithm has not displayed particular sensitivity.

\subsection{Model Selection}
\label{subsec:selection}

The best fitting covariance structure and number of components, is selected using an approximation of the Bayesian Information Criterion (BIC) \citep{schwarz78, kass95}. The BIC cannot be evaluated for clustMD models since the observed likelihood relies on the calculation of intractable integrals. However, the observed likelihood may be estimated as follows. The observed data vector $\underline{y}_i = (\underline{y}_i^\alpha, \underline{y}_i^\beta)$ where $ \underline{y}_i^\alpha = \underline{z}_i^\alpha \sim \sum_{g=1}^{G} \pi_g \mbox{MVN}(\underline{\mu}_g^\alpha, \Sigma_g^\alpha)$ and $\underline{y}_i^\beta \sim \mbox{Multinomial}(1, \underline{q})$. Treating these random variables as independent, the joint density can be approximated as the product of their marginals:
\begin{eqnarray}f(\underline{y}_i) \approx \left[\sum_{g=1}^{G} \pi_g \mbox{MVN}(\underline{z}_i^\alpha|\underline{\mu}_g^\alpha, \Sigma_g^\alpha)\right]\left[\prod_{m=1}^{M} q_{m}^{y^\beta_{im}}\right] \label{eqn:likeapprox}  \end{eqnarray}
The first term in (\ref{eqn:likeapprox}) is easily evaluated but the second term requires the probability of the observed categorical response pattern for observation $i$. i.e.
\begin{eqnarray}
&\hspace{-0.7cm}&q_m = \int_{\Omega_m}\sum_g \pi_g \mbox{MVN}(\underline{\mu}^\beta_g, \Sigma^\beta_g)d\underline{z}^\beta_i = \sum_g \pi_g \int_{\Omega_m} \mbox{MVN}(\underline{\mu}^\beta_g, \Sigma^\beta_g)d\underline{z}^\beta_i \nonumber \\
&\hspace{-0.7cm}&= \sum_g \pi_g \left[\prod_{j=C+1}^{O} \int_{\Omega_{mj}} N(\mu_{gj}, \sigma^2_{gj}) dz_{ij} \right] \left[\prod_{j=C+O+1}^{J} \int_{\Omega_{mj}} MVN(\underline{\mu}_{gj}, \Sigma_{gj})d\underline{z}_i^j \right] \label{eqn:pm}
\end{eqnarray}
The products in (\ref{eqn:pm}) consist of probabilities which were estimated in order to calculate $\tau_{ig}$ during the model fitting process. The products in the first term  in (\ref{eqn:pm}) are easily obtained from a normal distribution while the probabilities in the second are obtained by the Monte Carlo approximation described in Section~\ref{subsubsec:estep}. It should be noted that $q_m$ need only be estimated for the observed response patterns and not all $M$ possible response patterns. Thus the observed likelihood is approximated by:
$$\hat{\mathcal{L}} = \prod_{i=1}^{N} \left[\sum_{g=1}^{G} \pi_g \mbox{MVN}(\underline{z}_i^\alpha|\underline{\mu}_g, \Sigma_g)\right]\left[\prod_{m=1}^{M} \hat{q}_{m}^{y^\beta_{im}}\right]$$
The approximated BIC is then $\widehat{BIC} = 2\hat{\mathcal{L}} - \nu \log(N)$ where $\nu$ is the number of free parameters in the model. This approximation has been found to perform well, as illustrated through the simulation study detailed in Section~\ref{subsec:simdata}.

\section{Applications of the clustMD model}
\label{sec:applications}

The clustMD model was applied to simulated data and to a group of prostate cancer patients on whom mixed data have been recorded. Both contain continuous, ordinal and nominal variables. 
An application to another mixed data set is detailed in \cite{mcparland15}.

\subsection{Simulation Study}
\label{subsec:simdata}

One hundred data sets consisting of 800 observations of 10 variables was generated from a 2-cluster $VII$ model. Four of the variables were continuous, 3 were ordinal (with 2, 4 and 3 levels) and 3 were nominal (with 3, 3, and 4 levels). The categorical variables were obtained by transforming continuous values into categorical values in the manner described in Sections~\ref{subsec:Ord} and \ref{subsec:Nom}.

All 6 clustMD models were fitted to each data set with $G=1, \ldots, 4$. Each model was fitted using an MCEM algorithm with 1500 iterations. This was more than enough to reach convergence which was assessed as outlined in Section~\ref{subsubsec:mstep}. Examples of plots used to assess convegence are give in \cite{mcparland15}. To fit all 24 models in parallel took approximately 4.75 hours per data set. The algorithm was seeded with different initial partitions as outlined in Section~\ref{subsubsec:mstep} and there were very few differences in the results obtained from each of these initialisations.

\begin{figure}
\centering
  \includegraphics[height = 4cm,width=0.55\textwidth]{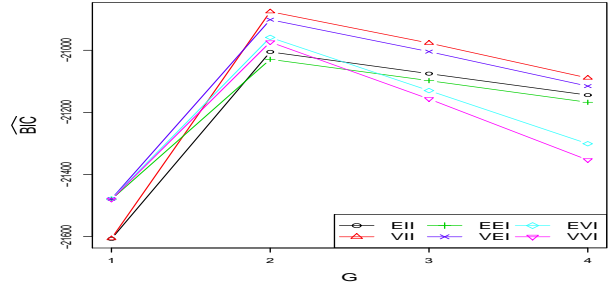}
\caption{\label{fig:SimBIC}Line plot of the estimated BIC values for each of the fitted models for the simulated data. The criterion was maximised by the $VII$ model with two clusters.}%
\label{fig:1}       
\end{figure}

The approximated BIC was evaluated for each of the 24 models fitted on each data set and the VII model with 2 components was chosen 96\% of the time. The median approximated BIC for each model and each value of $G$ is plotted in Figure~\ref{fig:SimBIC}. If the true cluster labels are compared to those identified by the clustMD model, it can be seen that the model performs very well. The mean adjusted Rand index (based on the optimal model using the approximated BIC) across the 100 data sets is 0.84.

If the above experiment is repeated, but where the data are generated from a 2-cluster model with non-diagonal covariance matrices, the approximated BIC tends to favour models with a larger number of components. These extra components allow the model to match the shape of the clusters that cannot be accommodated using diagonal models.


\subsection{Prostate Cancer Data}
\label{subsec:prostate}

This data set was analysed by \cite{byar80} and subsequently by \cite{hunt99}. The data may be found in \cite{andrews85}. Twelve mixed type measurements are available for 475 prostate cancer patients who were diagnosed as having either stage 3 or 4 prostate cancer. Of the variables, 8 are continuous, 3 are ordinal and 1 is nominal. The variables analysed are presented in Table~\ref{tab:vars}. The number of categories in the nominal `electrocardiogram' variable was reduced to 3 by combining categories since the Monte Carlo approximation can be inefficient when there is a small number of observations in a particular category for a particular cluster. Some other variables were recorded, such as the post trial survival status of the patients and the cause of death of those patients who died over the course of the trial.

\begin{table}
\caption{Variables analysed in the prostate cancer data set. The type of variable is denoted by a letter: continuous (C), ordinal (O) and nominal (N). The number in parentheses after the categorical variables indicates the number of possible responses for that variable.}
\label{tab:vars}
\centering
\begin{tabular}{llll}
\hline\noalign{\smallskip}
  Variable 			& Type 	&  Variable 					& Type\\
\noalign{\smallskip}\hline\noalign{\smallskip}
  Age 				& C	&  Index of tumour stage and histolic grade  	& C \\ 
  Weight 			& C	&  Serum prostatic acid phosphatase 		& C \\ 
  Systolic blood pressure 	& C 	&  Performance rating  				& O (4)\\ 
  Diastolic blood pressure 	& C 	&  Cardiovascular disease history 		& O (2)\\ 
  Serum haemoglobin 		& C	&  Bone metastasis 				& O (2)\\ 
  Size of primary tumour 	& C 	&  Electrocardiogram code 			& N (3)\\ 
\noalign{\smallskip}\hline
\end{tabular}
\end{table}

\begin{figure}
\begin{tabular}{lr}
\subfigure[Line plot of the $\widehat{BIC}$ values for each of the fitted models for the prostate cancer data. Best model is $EVI$ with $G=3$.]{\includegraphics[height = 4cm,width=0.44\textwidth]{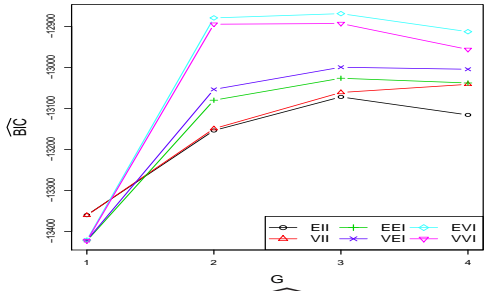} \label{fig:ByarBIC}}
&
\subfigure[Values of each dimension of $\hat{\underline{\mu}}_g$ across the 3 clusters. The categorical variables are labelled with capital letters.]{
\includegraphics[height = 4cm,width=0.52\textwidth]{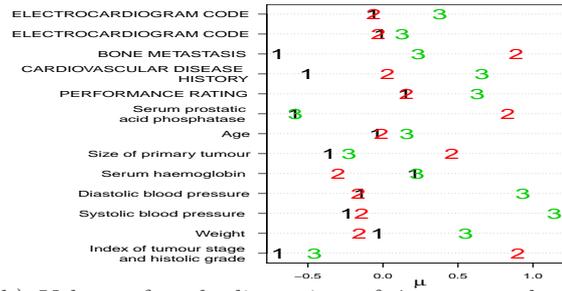}\label{fig:ByarMean}}
\end{tabular}
\caption{Line plot of $\widehat{BIC}$ values and plot of estimated group mean parameters.}
\label{fig:ByarBICMean}
\end{figure}

The suite of 6 clustMD models were fitted to the set of prostate cancer patients with the number of clusters ranging from 1 to 4. Fitting these 24 models in parallel using a MCEM algorithm with 3000 iterations took approximately 21 hours. No effort has been made to optimise the code and 3000 iterations are more than was required to reach convergence for all models. A line plot of the approximated BIC values is presented in Figure~\ref{fig:ByarBIC}. The model which maximises this criterion is a 3 cluster model, with the $EVI$ covariance structure.

\cite{hunt99} sought to identify the cancer stage and only considered 2 cluster models. It is interesting that a 3 cluster model is chosen here. A cross tabulation of the cluster labels versus the cancer stage diagnosis is given in Table~\ref{tab:byarClusG3}. It seems reasonable to expect clusters 1 and 3 to be similar given that both clusters consist primarily of stage 3 patients. However, comparing the mean vectors for these clusters (Figure~\ref{fig:ByarMean}), it can be seen that patients in cluster 3 are on average heavier and have higher levels of blood pressure. They are more likely to have a history of cardiovascular disease and their electrocardiogram score is more likely to indicate a serious anomaly. This suggests that a cardiovascular health issue differentiates patients in cluster 3 from those in cluster 1. Indeed, by examining the post trial survival status it can be seen that 21\% of cluster 3 patients are alive at the end of the trial, only 7\% died of prostatic cancer but 51\% died from heart or vascular disease or a cerebrovascular accident (stroke). The remaining 21\% died from other causes.

\begin{table}
\caption{Cross tabulation of estimated cluster labels versus the diagnosed prostate cancer stage. The adjusted Rand index is 0.49.}
\label{tab:byarClusG3}
\centering
\begin{tabular}{lll}
\hline\noalign{\smallskip}
 & Stage 3 & Stage 4 \\ 
\noalign{\smallskip}\hline\noalign{\smallskip}
  Cluster 1 & 207 &  14 \\ 
  Cluster 2 &  21 & 175 \\ 
  Cluster 3 &  45 & 13 \\
\noalign{\smallskip}\hline
\end{tabular}
\end{table}
Figure~\ref{fig:ByarMean} also shows that patients in cluster 2 have, on average, larger tumours and higher levels of serum prostatic acid phosphatase than patients in clusters 1 and 3. They are also more likely to have bone metastases. Analysing the survival status of clusters 1 and 2 it can be seen that 49\% of patients in cluster 2 died from prostatic cancer (as compared to 10\% in cluster 1), 21\% survived until the end of the trial (37\% in cluster 1) and 30\% died from other causes (57\% in cluster 1). 

\section{Discussion}
\label{sec:discussion}

The clustMD model presented here provides a suite of parsimonious mixture models for clustering mixed type data. The latent variable framework provides an elegant unifying structure for clustering this type of data.

Possible future research directions are plentiful. The most obvious deficiency of the proposed clustMD model is the local independence assumption. It would be very beneficial to allow for full covariance matrices that can model dependencies between mixed type variables. An intermediate step to achieving this goal is to consider a block diagonal covariance matrix which would allow for dependencies between variables of the same type. The associated eigenvalue decomposition of such a covariance matrix would then need to be considered.

The Monte Carlo approximation used in the E-step of the model fitting algorithm is a simple and effective solution but it is not without issues. If the probability of observing a particular response on a nominal variable is very small for a particular cluster then a large number of Monte Carlo samples may be required to observe a response in this category. This can slow the model fitting algorithm or even cause instability. A more efficient way to approximate the intractable integrals could improve the model fitting efficiency.

Unless a very large number of Monte Carlo simulations are required for the E-step, due to a sparsely observed nominal variable, the model fitting algorithm is very computationally efficient. Indeed, if no nominal variables are present each model is fitted in a matter of seconds. The suite of clustMD models are available through the R package {\tt clustMD} \citep{R2013}.


Finally, the mixture model proposed here is assumed to consist of Gaussian distributions but this need not be the case. Mixtures of other distributions could be used instead. Heavier tailed distributions such at the $t$ distribution or the normal inverse Gaussian distribution may allow for more extreme observations in both the continuous and categorical variables. This extension could potentially draw on the work of \cite{mclachlan98, karlis09} and \cite{ohagan12b}.

\section*{Acknowledgements}
The authors would like to thank Science Foundation Ireland, grant number 09/RFP/MTH2367, for supporting this research.
\bibliographystyle{plainnat}
\bibliography{total.bib}{}   

\end{document}